\documentclass{article}

\usepackage{PRIMEarxiv}

\usepackage[utf8]{inputenc} 
\usepackage[T1]{fontenc}    
\usepackage{hyperref}       
\usepackage{url}            
\usepackage{booktabs}       
\usepackage{amsfonts}       
\usepackage{nicefrac}       
\usepackage{microtype}      
\usepackage{lipsum}
\usepackage{fancyhdr}       
\usepackage{graphicx}       
\graphicspath{{media/}}     
\usepackage{amsmath} 
\usepackage[numbers]{natbib}
\usepackage{multirow}
\usepackage[table]{xcolor}
\usepackage{appendix} 
\usepackage{chngcntr} 
\usepackage{pdfpages}
\usepackage{setspace}
\usepackage{threeparttable}
\usepackage{caption}

\usepackage{soul}
\colorlet{lgreen}{green!40}
\colorlet{lorange}{orange!40}

\begin{document}

\title{Deep Learning Pipeline for Fully Automated Myocardial Infarct Segmentation from Clinical Cardiac MR Scans}

\author{
  Matthias Schwab, Mathias Pamminger, Christian Kremser, Agnes Mayr \\
  Department of Radiology \\
  Medical University of Innsbruck \\
  Innsbruck\\
   \And
  Markus Haltmeier \\
  Department of Mathematics \\
  University of Innsbruck \\
  Innsbruck\\
}

\maketitle

\begin{abstract}
\textbf{Purpose:} To develop and evaluate a deep learning-based method that allows to perform myocardial infarct segmentation in a fully-automated way.

\textbf{Materials and Methods:} For this retrospective study, a cascaded framework of two and three-dimensional convolutional neural networks (CNNs), specialized on identifying ischemic myocardial scars on late gadolinium enhancement (LGE) cardiac magnetic resonance (CMR) images, was trained on an in-house training dataset consisting of $144$ examinations. On a separate test dataset from the same institution, including images from $152$ examinations obtained between 2021 and 2023, a quantitative comparison between artificial intelligence (AI)-based segmentations and manual segmentations was performed. Further, qualitative assessment of segmentation accuracy was evaluated for both human and AI-generated contours by two CMR experts in a blinded experiment.

\textbf{Results:}
Excellent agreement could be found between manually and automatically calculated infarct volumes ($\rho_c = 0.9$). The qualitative evaluation showed that compared to human-based measurements, the experts rated the AI-based segmentations to better represent the actual extent of infarction significantly ($p < 0.001$) more often ($33.4 \%$ AI, $25.1 \%$ human, $41.5 \%$ equal). On the contrary, for segmentation of microvascular obstruction (MVO), manual measurements were still preferred ($11.3\%$ AI, $55.6\%$ human, $33.1 \%$ equal).

\textbf{Conclusion:} This fully-automated segmentation pipeline enables CMR infarct size to be calculated in a very short time and without requiring any pre-processing of the input images while matching the segmentation quality of trained human observers. In a blinded experiment, experts preferred automated infarct segmentations more often than manual segmentations, paving the way for a potential clinical application.

\end{abstract}


\section{Introduction}

Ischemic heart disease remains a leading cause of global mortality, responsible for approximately 9.1 million deaths worldwide in 2019  \citep{nowbar2019mortality, vos2020global}. It has been shown that following ST-segment elevation myocardial infarction, accurately assessing infarct size and microvascular obstruction (MVO) are crucial for clinical decision-making and for prediction of major adverse cardiovascular events \cite{kim2000use, gerber2012prognostic, larose2010predicting, de2017relationship}. However, obtaining these important predictors requires segmentation of late gadolinium enhancement (LGE) cardiac magnetic resonance (CMR) images. 

As manual LGE segmentation by expert readers is time-consuming and additionally yields limited reproducibility \citep{flett2011evaluation}, recently a lot of work has been done developing deep learning-based algorithms for automatic infarct segmentation \cite{moccia2019development,fahmy2020three,yue2019cardiac,brahim2022improved,popescu2022anatomically,zabihollahy2020fully,chen2022automatic,al2024deep}. The topic received even more attention when two challenges focusing on myocardial infarct segmentation were held in the course of the 2020 MICCAI conference \cite{lalande2022deep,li2023myops}.  However, a lot of these frameworks still have major drawbacks, such as ignoring the extensive image preprocessing steps that would be necessary when applied in clinical practice. Furthermore, the segmentation performance was only measured quantitatively by comparing with human-created ground truth measurements. Recent findings call into question if metrics such as the Dice coefficient can be accepted as the \textit{de facto gold standard} for measuring segmentation quality beyond expert opinion \cite{kofler2023we}. Therefore, to be able to develop clinically helpful segmentation models, a better understanding of the subjective quality perception of clinical experts is required \cite{hoebel2023expert}. Although there are some methods in the literature for qualitatively assessing segmentation accuracy \cite{mcguinness2010comparative,shi2014jaccard}, these often do not provide information about the specific types of segmentation errors and their potential effects in a clinical setting. This means that in medical image segmentation, subjective performance evaluation heavily depends on the underlying medical application and includes diverse approaches. These involve measuring time that experts need to manually correct automatically generated segmentations \cite{lu2021randomized, wang2018interactive}, rating the segmentation quality \cite{di2021application, mitchell2020deep}, or blindly comparing manual ground truth and automatic segmentations \cite{mitchell2020deep}. However, to the best knowledge of the authors, no qualitative assessment for artificial intelligence (AI)-generated myocardial infarct segmentation has yet been published. 

The purpose of this study was to develop and evaluate a deep learning-based algorithm that enables accurate and fast segmentation of myocardial infarction and MVO on clinical LGE CMR images. The developed pipeline allows to quantify the extent of myocardial infarction on clinical LGE CMR images in a fully automated way. This is done without any human intervention, i.e. the preprocessing steps required for accurate CNN segmentation of the clinical data are also fully automated. To validate the segmentation performance of the developed framework, not only the usual quantitative metric between human-created ground truth measurements were calculated. Additionally, a comprehensive qualitative evaluation study incorporating the experience and knowledge of two CMR-specialized and certified radiologists was carried out. 

\section{Materials and Methods}
\label{sec:methods}

This study is concerned with the retrospective analysis of quantitative and qualitative performance of a deep learning segmentation algorithm for myocardial infarct quantification on clinical data. All the LGE CMR images that were used in our study were originally acquired prospectively as part of the MARINA-STEMI (Magnetic Resonance Imaging In Acute ST-Elevation Myocardial Infarction) study (NCT04113356), which was approved by the local ethics committee, with all patients providing written informed consent prior to inclusion. For both development and testing of the segmentation algorithm, a total of $329$ examinations were randomly selected from the MARINA-STEMI cohort and assigned to either the training, evaluation, or test datasets. While several articles have been published in the last decade \cite{MAYR20221030,lechner2022impact,lechner2022association} using patients from this cohort to address clinical questions, this paper is the first to take a machine learning approach to these data.

\subsection{Training Dataset}
For training of the algorithm, an in-house training dataset consisting of 144 LGE CMR examinations from 142 unique patients (baseline: $n = 54$; 4 months follow-up: $n = 24$; 12 months follow-up: $n = 66$) was created from data collected at the Department of Radiology. During training, segmentation performance was evaluated after each epoch on a hold-out evaluation dataset consisting of $33$ LGE examinations (Table \ref{tab:demograph}). Manual segmentations of the left ventricle (LV) were done by medical experts using the local routine diagnostic interpretation and reporting software (DeepUnity Diagnost, Dedalus Healthcare Systems Group, Germany) according to the guidelines explained in Appendix \ref{sec:S2} and Fig \ref{fig:S1}. In each of the short axis slices of the CMR, the following four tissue regions were segmented if present: remote myocardium, LGE-enhanced myocardium, MVO, and blood pool. Binary segmentation masks were created from these manually defined regions for training of the deep learning models. Since we were interested in a segmentation framework that is able to handle unprocessed MR images as they occur in clinical practice, we presented the complete LGE image stack, including LV outflow tract, apex, and slices without LV myocardium, to the CNN. In addition, close attention was paid to marking blood within the LV outflow tract. 

\subsection{Test Dataset}
The study analyzes the algorithm’s performance on a CMR LGE test dataset consisting of images obtained between 2021 and 2023 at the same institution as the training dataset. In total, images from $152$ LGE CMR measurements, including data from $121$ unique ST-segment elevation myocardial infarction patients after successful primary percutaneous coronary interventions (p-PCI), are analyzed. This includes images obtained within one week after p-PCI (baseline), 4 months, and 12 months follow-up examinations, respectively. The details of the dataset demographics are displayed in Table \ref{tab:demograph}. LGE CMR images were acquired on a 1.5 Tesla MR scanner (Magnetom AvantoFit, Siemens, Erlangen, Germany) 10 to 20 minutes after an intravenous gadolinium bolus injection of $0.2$ mmol/kg body mass (Gadobutrol, Gadovist, Bayer AG, Germany) using an ECG-triggered phase-sensitive inversion recovery sequence. Exact details on the dataset specifics as well as imaging protocols can be found in Appendix \ref{sec:S1}. Segmentation masks for LGE-enhanced myocardium as well as for MVO were drawn by appropriately trained scientific staff members of the study team under the supervision of CMR-specialized radiologists (Appendix \ref{sec:S2} and Fig \ref{fig:S1}).

\begin{table*}[t]
\caption{Patient demographics of the different datasets: Data are numbers of patients or means $\pm$ standard deviations, with ranges in parentheses.}
\centering
\begin{tabular}{lllll}
\toprule
{}                        & \textbf{Training}        & \textbf{Evaluation}     & \textbf{Test} \\
\midrule
\textbf{Number}           & $144$                    & $33$                    & $152$ \\
\textbf{Baseline}         & $54$                     & $12$                    & $33$ \\
\textbf{4FU}              & $24$                     & $3$                     & $43$ \\
\textbf{12FU}             & $66$                     & $18$                    & $76$ \\
\textbf{Age}              & $57 \pm 12 \; (29-84)$   & $59 \pm 13 \; (34-88)$  & $61 \pm 10 \; (42-86)$ \\
\textbf{Sex (m/f)}        & $118/26$                 & $24/9$                  & $126/16$  \\
\midrule       
\multicolumn{4}{l}{\footnotesize Baseline: Examination within one week after p-PCI } \\
\multicolumn{4}{l}{\footnotesize 4FU: Examination 4 months after p-PCI }\\
\multicolumn{4}{l}{\footnotesize 12FU: Examination 12 months after p-PCI }
\end{tabular} 
\label{tab:demograph}
\end{table*}

\subsection{AI Framework Development}

\subsubsection{Deep Learning Pipeline}
In summary, our deep learning pipeline consists of two main steps:

\begin{enumerate}
    \item Extracting a stack with smaller image sizes out of the original data that still contains the entire LV.
    \item Performing multiclass segmentation with special focus on the myocardial scar on the extracted volumes.
\end{enumerate}

\begin{figure}[b]
    \centering
    \includegraphics[width= \textwidth]{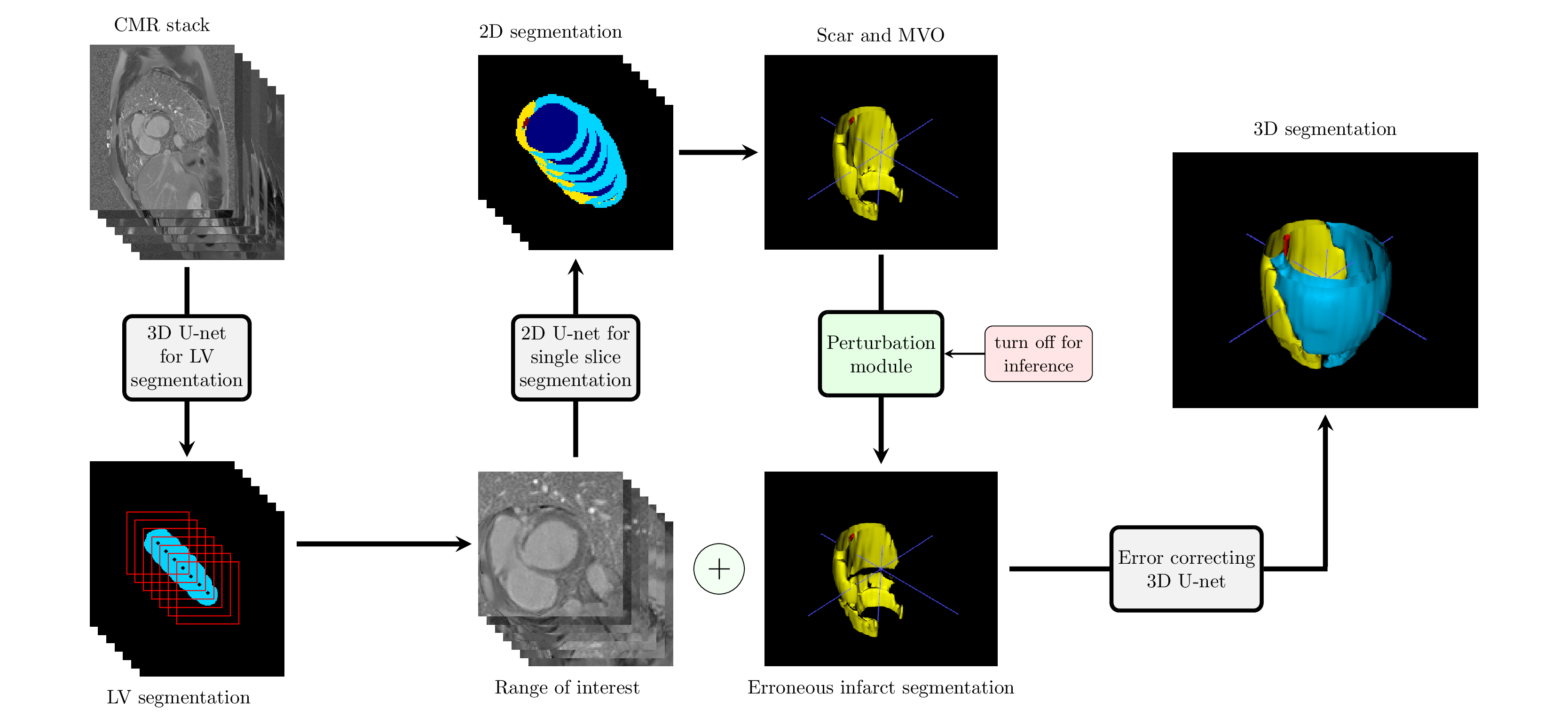}
    \caption{AI pipeline. Firstly, a smaller image stack is extracted out of the original data by segmentation of the left ventricle. Then an error correcting 2D-3D cascaded framework is used to perform multiclass segmentation on the left ventricle.}
    \label{fig:method}
\end{figure}
The overall configuration of the proposed deep learning framework is illustrated in Figure \ref{fig:method}. Since myocardial scar and MVO are potentially very small areas, it is a very hard task to segment them from the original CMR images as they are obtained in clinical practice. Therefore, in a preprocessing step, our framework is extracting a range of interest stack with smaller image sizes out of the original data, which still contains the entire LV. To this end, a 3D U-net was trained to segment the LV in the original image stack. The detailed architecture of the network and a description of how the model was trained can be found in Fig \ref{fig:S2} and Appendix \ref{sec:S3} of the supplemental material. After detection of the LV, the network’s segmentation of the middle slice of the CMR volume is extracted. Then the center of mass of the segmentation mask is calculated, obtaining the ”center“ of the LV. Finally, the smaller image stack is generated by excision of smaller images around the calculated center in all the slices.

In the second step, an error correcting 2D-3D cascaded network \cite{schwab2023error} was trained on the extracted image stacks. This method was specifically developed for infarct segmentation and tries to combine the strengths of both 2D and 3D architectures to be able to optimally segment the often thin and poorly contrasted myocardial scars. In short, the method works by first training on two dimensional images, utilizing the advantage that there are much more 2D images than 3D volumes available for training. After that, the three-dimensional CNN exploits both the relationships between the slices as well as preceding 2D segmentations. By introducing various perturbations to the 2D segmentation masks during training of the 3D network, an error correction characteristic is enforced for the 3D network, which in turn leads to a better performance of the method on new unseen data. Details about the used architectures, perturbations, and training strategies can be found in the supplemental material (Appendix \ref{sec:S3}, \ref{sec:S4} and Fig \ref{fig:S2}, \ref{fig:S3}).

\subsubsection{Evaluation of Segmentation Performance}

Our evaluation of segmentation performance analysis consists of:
\begin{enumerate}
    \item A quantitative assessment of segmentation accuracy comparing AI-segmentations to manual markings.
    \item A qualitative assessment of segmentation accuracy done by CMR experts.
\end{enumerate}

For the qualitative assessment of segmentation accuracy both, manual and automatic segmentation masks were evaluated by two CMR experts in a blinded experiment. For each patient, we distributed manual and automatic segmentations randomly into \textit{segmentation A} and \textit{segmentation B}. Not knowing which mask was created by humans and which by AI, the medical experts had to subjectively assess the segmentation quality of LGE and MVO segmentations. On a per-slice level, they had to decide for each segmentation between different ratings (Fig \ref{fig:S4}):
\begin{itemize}
    \item \textbf{optimal}: The segmentation was done to their full satisfaction.
    \item \textbf{too big}: An infarct/MVO was correctly identified. However, the area marked was too big.
    \item \textbf{too small}: An infarct/MVO was correctly detected. However, too small an area was marked. 
    \item \textbf{wrong organ}: Areas outside the heart were marked as infarct/MVO.
    \item \textbf{false negative}: An infarct/MVO was completely overlooked in this slice.
    \item \textbf{false positive}: An area in the myocardium was falsely marked as infarct/MVO in a slice where no infarct/MVO is present.
    \item \textbf{true negative}: Rightfully nothing was marked in a slice where no infarct/MVO is present.
\end{itemize}

After evaluating the segmentation individually, the experts additionally had to look at the two methods side by side and decide with which of the two segmentations they agreed more. For this task, they were able to choose between \textit{segmentation A}, \textit{segmentation B}, or \textit{equally good} (Fig \ref{fig:S5}).

The data was distributed between the two raters the following way: For a randomly chosen subset of 20 patients, both experts gave their ratings independently of each other. Then the agreement between their answers was evaluated, and cases where the experts disagreed were discussed in more detail. The remaining dataset was then split between the two experts so that each evaluated half of the remaining patients. For images in which the qualitative assessment was not entirely clear, the experts reached a consensual decision. Further details about the design of the qualitative experiments can be found in the supplemental material (Appendix \ref{sec:S5}).

\subsection{Statistical Analysis}

To quantify the segmentation accuracy of our method, we calculated different metrics between AI and human-generated measurements. These incorporate clinical as well as geometrical metrics. For assessing the geometrical agreement between the methods, Dice similarity coefficients (DICE) were calculated. Further, absolute volume difference (AVD) in $\mathrm{ml}$ as well as absolute volume difference rate (AVDR) with respect to the volume of the myocardium ($V_{\mathrm{MYO}}$) were calculated. The performance metrics between AI segmentations $P$ and manual segmentations $G$ were obtained as follows:
\begin{align}
    \label{eq:dice}
    \mathrm{DICE} &= \frac{2 |P \cap G|}{|P|+|G|}, \\
    \mathrm{AVD} &= ||P|-|G|| \times \text{voxelvolume}, \\
    \mathrm{AVDR}&= \frac{\mathrm{AVD}}{V_{\mathrm{MYO}}}, 
\end{align}
where $|\cdot|$ denotes the cardinality of a set.
    
To quantify the agreement and reliability between AI and human-based infarct size measurements, we calculated the infarct volumes as a percentage of the total LV myocardial mass for both methods. Statistical analysis included a paired Wilcoxon signed rank test, Lin's concordance correlation coefficient ($\rho_c$) \cite{lawrence1989concordance}, and Bland-Altman analysis. 

In the qualitative analysis, we compared relative proportions of the given answers for both human-based and AI-based segmentations. In order to identify significant differences in frequency between the experts' assessments, a one-way chi-square test was used. Further, rater agreement was investigated by calculating confusion matrices and Cohen’s kappa coefficients ($\kappa$). 

Statistical analysis was performed with Python (version 3.9) using the scipy package. For all statistical tests, a statistical significance threshold of $0.05$ was used.

\section{Results}
\label{sec:results}

\subsection{Quantitative Segmentation Accuracy}

\begin{figure}[!b]
  \centering
  \includegraphics[width= \textwidth]{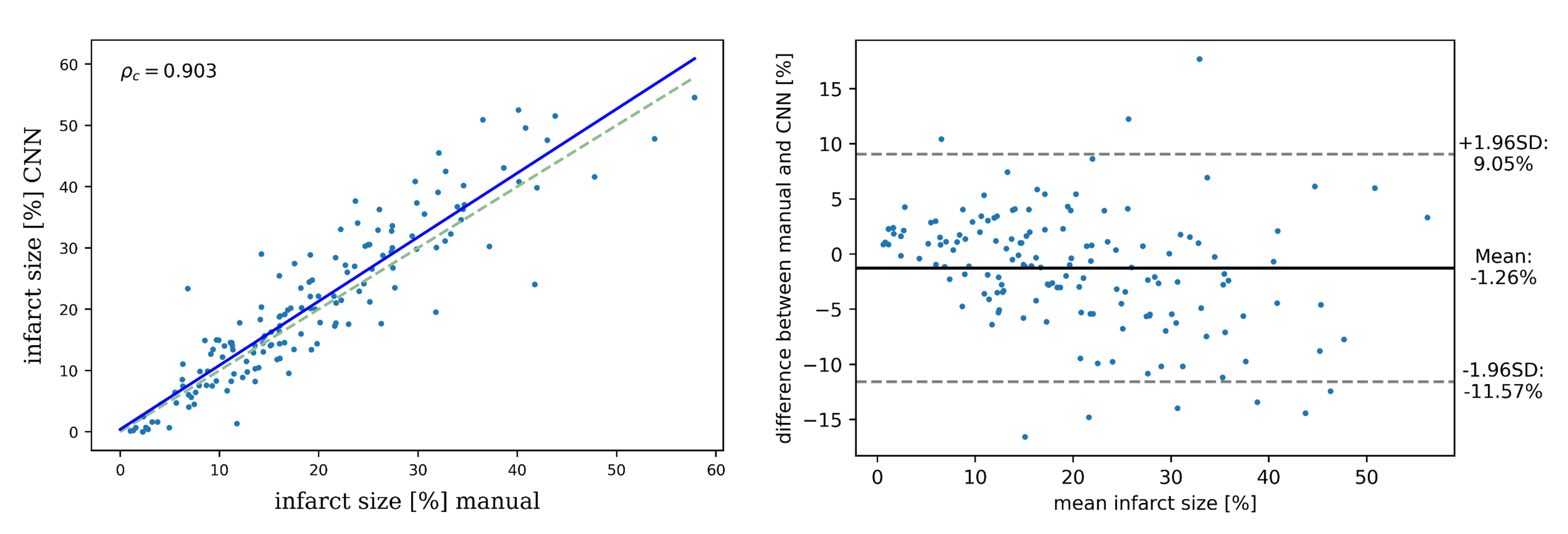}
  \caption{Scatter plot (left) and Bland-Altman analysis (right) of infarct size as a percentage of the total myocardial volume determined automatically and manually. In the scatter plot, the dashed line indicates $100$ percent agreement, and the solid line represents the linear regression line.}
  \label{fig:volumes}
\end{figure}

The deep learning method reached mean Dice coefficients of $64.11 \%$ for infarct segmentation and $82.20 \%$ for MVO segmentation. The mean AVD between manually and automatically calculated infarct volumes was $4.97 \mathrm{ml}$, and the mean AVDR was $4.04 \%$. For MVO, only very small volume differences were found, with a mean AVD of $0.59 \mathrm{ml}$ and a mean AVDR of $0.43 \%$. However, only in $15 \%$ of all the patients in the test dataset MVO was present As the Dice coefficient is undefined when both the ground truth and the predicted segmentation masks are empty, it was set to $1$ for such cases. This resulted in optimal metric values for all the patients where the method correctly detected no MVO.  When only taking into account patients where MVO was present, the mean Dice score reduced significantly to $25.04 \%$, and also for AVD and AVDR, the accuracy decreased considerably (see Table \ref{tab:quant}). 

\begin{table}[t]
 \caption{Different quantitative metrics for infarct and MVO segmentation on the $152$ examinations of the test dataset. The right-hand column contains the MVO results obtained exclusively for those patients in whom MVO was present ($n = 23$).}
  \centering
  \begin{tabular}{l|c|c|c}
                                    & Infarction        & MVO               & MVO only     \\
    \midrule
    $\mathrm{DICE} (\%)$            & $64.11 \pm 18.37$ & $85.20 \pm 32.93$ & $25.04 \pm 25.28$    \\
    $\mathrm{AVD} (\mathrm{ml})$    & $4.97 \pm 5.07$   & $0.59 \pm 2.14$   & $3.92 \pm 4.39$ \\
    $\mathrm{AVDR} (\%)$            & $4.04 \pm 3.5 $   & $0.43 \pm 1.49$   & $2.86 \pm 2.93$  \\
    \midrule
    \multicolumn{4}{l}{\footnotesize DICE: Dice similarity coefficient }\\
    \multicolumn{4}{l}{\footnotesize AVD: Absolute volume difference }\\
    \multicolumn{4}{l}{\footnotesize AVDR: absolute volume difference rate w.r.t. myocardium volume.}
  \end{tabular}  
  \label{tab:quant}
\end{table}

Scatter plot and Bland-Altman analysis showed good agreement between manually and automatically calculated infarct sizes, expressed as a percentage of the total myocardial volume (Figure \ref{fig:volumes}). Concordance correlation was very high ($\rho_c = 0.903$, $95 \%$ CI $[0.871, 0.923]$), and Bland-Altman analysis showed little average difference of $-1.26 \%$ between manual and CNN volume calculations. However, the bias of the neural network to mark slightly bigger scars is statistically significant ($p < 0.01$). The limits of agreement in Bland-Altman analysis ranged from $-11.57 \%$ to $9.05 \%$.

\subsection{Qualitative Segmentation Performance}

\begin{figure}[t]
    \centering
    \includegraphics[width=0.90\textwidth]{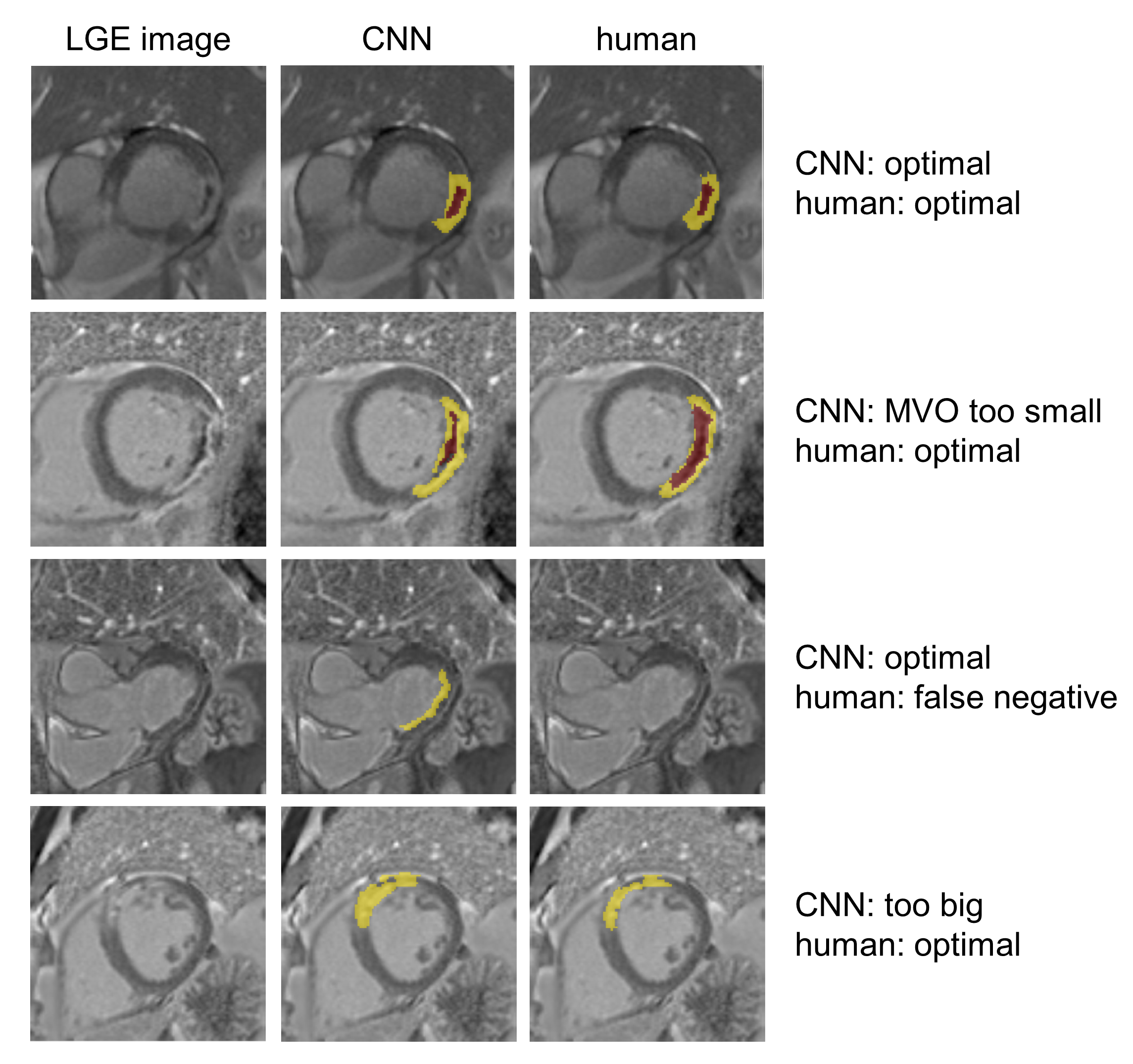}
    \caption{Examples of infarct segmentations, which include both optimal and faulty CNN and human-based segmentations. Expert ratings for the corresponding images are displayed on the right-hand side.}
    \label{fig:examples}
\end{figure}

Subjective evaluation of the segmentation performance was done for both LGE segmentation and MVO segmentation on a per-slice level on the $152$ patients in the test dataset. This all together resulted in ratings for LGE and MVO segmentation on $1619$ pairs of CMR slices. Examples of automatically and manually created segmentation masks with corresponding expert ratings are displayed in Figure \ref{fig:examples}. For the evaluation of the direct comparison between AI and human-created segmentation masks MRI slices in which both methods correctly showed no scar/no MVO were excluded, as the segmentations in these slices could only be evaluated as equally good.

Based on the experts validation of the segmentations, we investigated the diagnostic performances of human-based and AI-based predictions. For myocardial scars, diagnostic performance was very high, as the AI framework only missed two scars in the whole dataset. However, these two scars were tiny and could only be clearly confirmed by the CMR experts after an additional review of the functional images and previous examinations (Fig \ref{fig:S6}). For MVO detection, though, AI-based predictions showed a considerably lower sensitivity ($65 \%$) compared to humans ($91 \%$). Contingency tables and corresponding sensitivity and specificity values are shown in Table ~\ref{tab:cont}.

\begin{table}[!b]
\fontsize{9pt}{11pt}\selectfont
\caption{Diagnostic performances of AI and humans on a per-patient level based on the experts blinded validations of the segmentations. Contingency tables for AI-based and human-based scar and MVO detection. Sensitivity and specificity levels are displayed with $95 \%$ confidence intervals.}
\label{tab:cont}
\begin{minipage}{.5\linewidth}
    \centering
    \begin{tabular}{c c p{1.1cm} p{1.1cm} c }
    
     \multicolumn{2}{l}{\cellcolor{olive!25}\textbf{Sensitivity:} $99\% \, [95.3,99.8]$}   & \multicolumn{2}{l}{\cellcolor[gray]{.9}ground truth based}  &  \\
     \multicolumn{2}{l}{\cellcolor{olive!25}\textbf{Specificity:} $-$}   & \multicolumn{2}{l}{\cellcolor[gray]{.9}on experts validation}  &  
       \vspace{0.2cm} \\

      &   & scar            & no scar                                                        & $\sum$ \\
     \cellcolor[gray]{.9} AI-based & scar & \cellcolor{blue!73} $150$  & \cellcolor{blue!5}$0$ & $150$ \\ 
     \cellcolor[gray]{.9}prediction & no scar & \cellcolor{blue!12}$2$ & \cellcolor{blue!5}$0$ & $2$ \\ 
     & $\sum$ & $152$ & $0$ & $152$ \\
    \midrule 
   \multicolumn{2}{l}{\cellcolor{olive!25}\textbf{Sensitivity:} $100\% \, [97.6,100]$}   & \multicolumn{2}{l}{\cellcolor[gray]{.9}ground truth based}  &  \\
    \multicolumn{2}{l}{\cellcolor{olive!25}\textbf{Specificity:} $-$} & \multicolumn{2}{l}{\cellcolor[gray]{.9}on experts validation}  &  
       \vspace{0.2cm} \\
    &   & scar            & no scar                                                        & $\sum$ \\
     \cellcolor[gray]{.9}human-based & scar & \cellcolor{blue!74}$152$  & \cellcolor{blue!5}$0$ & $152$ \\ 
     \cellcolor[gray]{.9}prediction & no scar & \cellcolor{blue!5}$0$ & \cellcolor{blue!5}$0$ & $0$ \\ 
     & $\sum$ & $152$ & $0$ & $152$ \\
    \end{tabular}
\end{minipage}
\begin{minipage}{.5\linewidth}
    \centering
    \begin{tabular}{c c p{1.1cm} p{1.1cm} c }
    
     \multicolumn{2}{l}{\cellcolor{olive!25}\textbf{Sensitivity:} $65\% \, [42.7,83.6]$}   & \multicolumn{2}{l}{\cellcolor[gray]{.9}ground truth based}  &  \\
     \multicolumn{2}{l}{\cellcolor{olive!25}\textbf{Specificity:} $97\% \,[92.3,99.2]$}   & \multicolumn{2}{l}{\cellcolor[gray]{.9}on experts validation}  &  
       \vspace{0.2cm} \\

      &   & MVO            & no MVO                                                        & $\sum$ \\
     \cellcolor[gray]{.9} & MVO & \cellcolor{blue!25} $15$  & \cellcolor{blue!15}$4$ & $19$ \\ 
     \multirow{-2}{1.88cm}{\cellcolor[gray]{.9}AI-based prediction} & no MVO & \cellcolor{blue!20}$8$ & \cellcolor{blue!65}$125$ & $132$ \\ 
     & $\sum$ & $23$ & $129$ & $152$ \\
    \midrule 
   \multicolumn{2}{l}{\cellcolor{olive!25}\textbf{Sensitivity:} $91 \% \, [72.0,99.0]$}   & \multicolumn{2}{l}{\cellcolor[gray]{.9}ground truth based}  &  \\
     \multicolumn{2}{l}{\cellcolor{olive!25}\textbf{Specificity:} $99 \% \, [95.8,100]$}   & \multicolumn{2}{l}{\cellcolor[gray]{.9}on experts validation}  &  
       \vspace{0.2cm} \\
    &   & MVO            & no MVO                                                        & $\sum$ \\
     \cellcolor[gray]{.9} & MVO & \cellcolor{blue!35}$21$  & \cellcolor{blue!10}$1$ & $22$ \\ 
     \multirow{-2}{1.88cm}{\cellcolor[gray]{.9}human-based prediction} & no MVO & \cellcolor{blue!12}$2$ & \cellcolor{blue!70}$128$ & $130$ \\ 
     & $\sum$ & $23$ & $129$ & $152$ \\
    \end{tabular}
\end{minipage}
\end{table}

\begin{figure}[t]
    \centering
    \includegraphics[width= \textwidth]{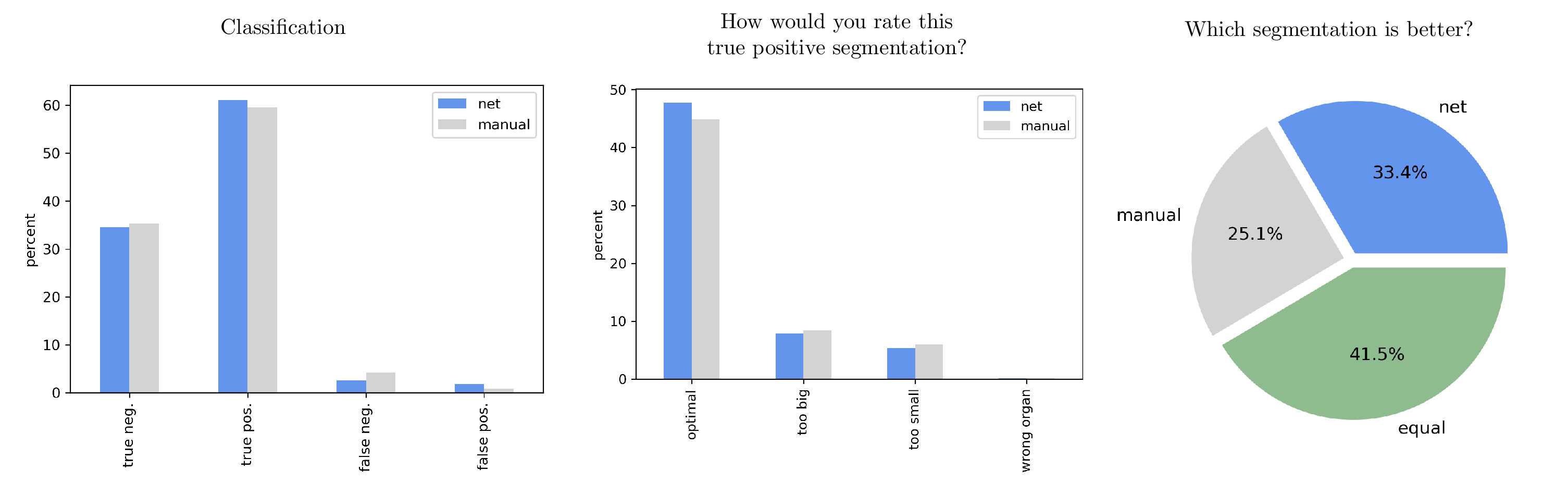}
    \caption{Experts rating for manual and automatic LGE segmentations. All slices were classified into true negative, true positive, false negative, and false positive (left). For true positive segmentations, the raters had to provide more detailed feedback (middle) and finally compare the two segmentation methods (right).}
    \label{fig:LGE}
\end{figure}

For LGE segmentation, raters overall preferred the automatic measurements, see Figure \ref{fig:LGE}. In $33.5 \%$ of the cases, they decided to agree more with the segmentation done by the neural net, whereas in only $25.1 \%$ of the cases manual segmentation was preferred. When excluding all cases that were rated as equal, a one-way chi-square test revealed that AI segmentations were preferred significantly more often ($p < 0.001$). On a per-slice level, the experts were fully satisfied (optimal segmentation or true negative) with the network’s performance in $82.2 \%$ of all the evaluated cases. This is slightly higher than for the human-based measurements, where experts expressed full agreement in $80.2 \%$ of cases. The main difference in performance was that fewer scars were overlooked (false negative) by the CNN ($2.6 \%$) compared to humans ($4.3 \%$). However, the fraction of wrongly marked infarct scars (false positive) was bigger for the CNN-generated segmentations ($1.8 \%$) compared to the human-created contours ($0.8 \%$). Total failure due to the marking of a myocardial scar in a wrong organ has hardly ever been observed with either method ($< 1 \%$).

\begin{figure}[t]
    \centering
    \includegraphics[width= \textwidth]{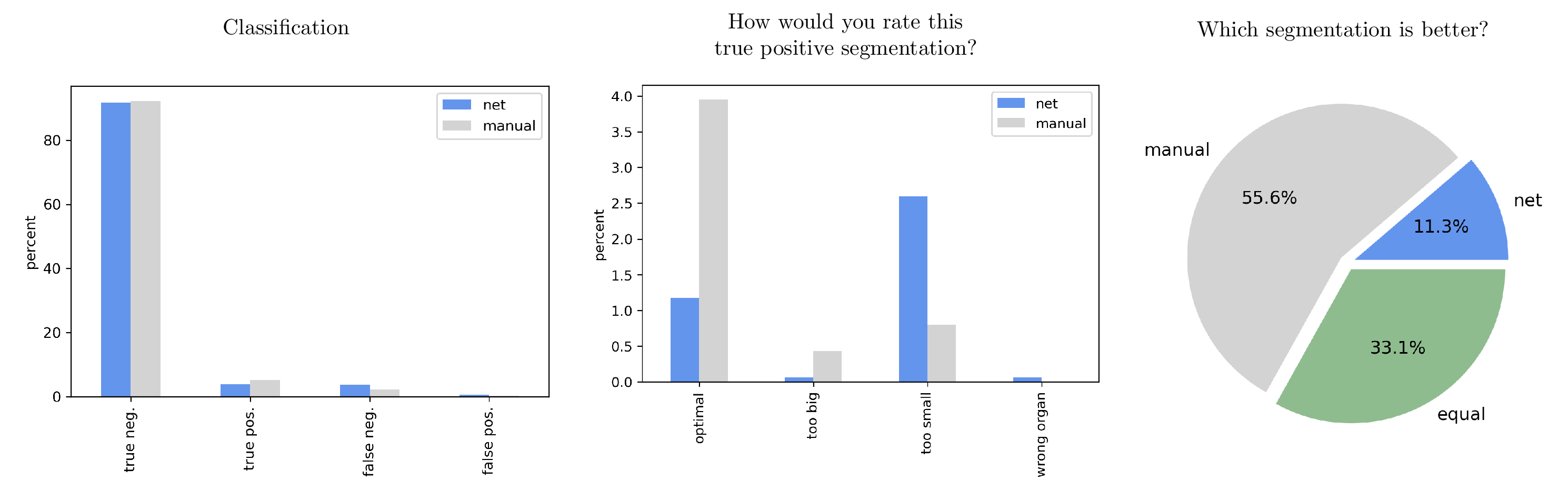}
    \caption{Experts rating for manual and automatic MVO segmentations. All slices were classified into true negative, true positive, false negative, and false positive (left). For true positive segmentations, the raters had to provide more detailed feedback (middle) and finally compare the two segmentation methods (right).}
    \label{fig:MVO}
\end{figure}

In contrast to LGE segmentation, for MVO quantification, manually created measurements were still superior to the deep learning algorithm, as illustrated in Figure \ref{fig:MVO}. The experts decided significantly more often ($p < 0.001$) in favor of the manual MVO segmentations ($55.6 \%$) compared to the CNN-generated measurements ($11.3 \%$). In the remaining $33.1 \%$ of cases, the two segmentations were rated as equally good. The main difference between manual and automatic segmentations was in sensitivity. CNN-generated segmentation missed MVO (false negative) in quite a few of the slices ($3.8 \%$), especially when considering that MVO was only present in $7.4 \%$ of all slices. Furthermore, also in slices where MVO was correctly detected, the framework tended to mark a too small area ($2.6 \%$), where in one slice MVO was marked in a wrong organ by the CNN. In contrast to that, $4.0 \%$ of human segmentations were rated as optimal, and only $0.8 \%$ were considered as too small. However, also humans missed a substantial number of slices (false negative) where MVO was present ($2.2 \%$).

\subsubsection{Rater Agreement}

\begin{figure}[t]
    \centering
    \includegraphics[width=0.95 \textwidth]{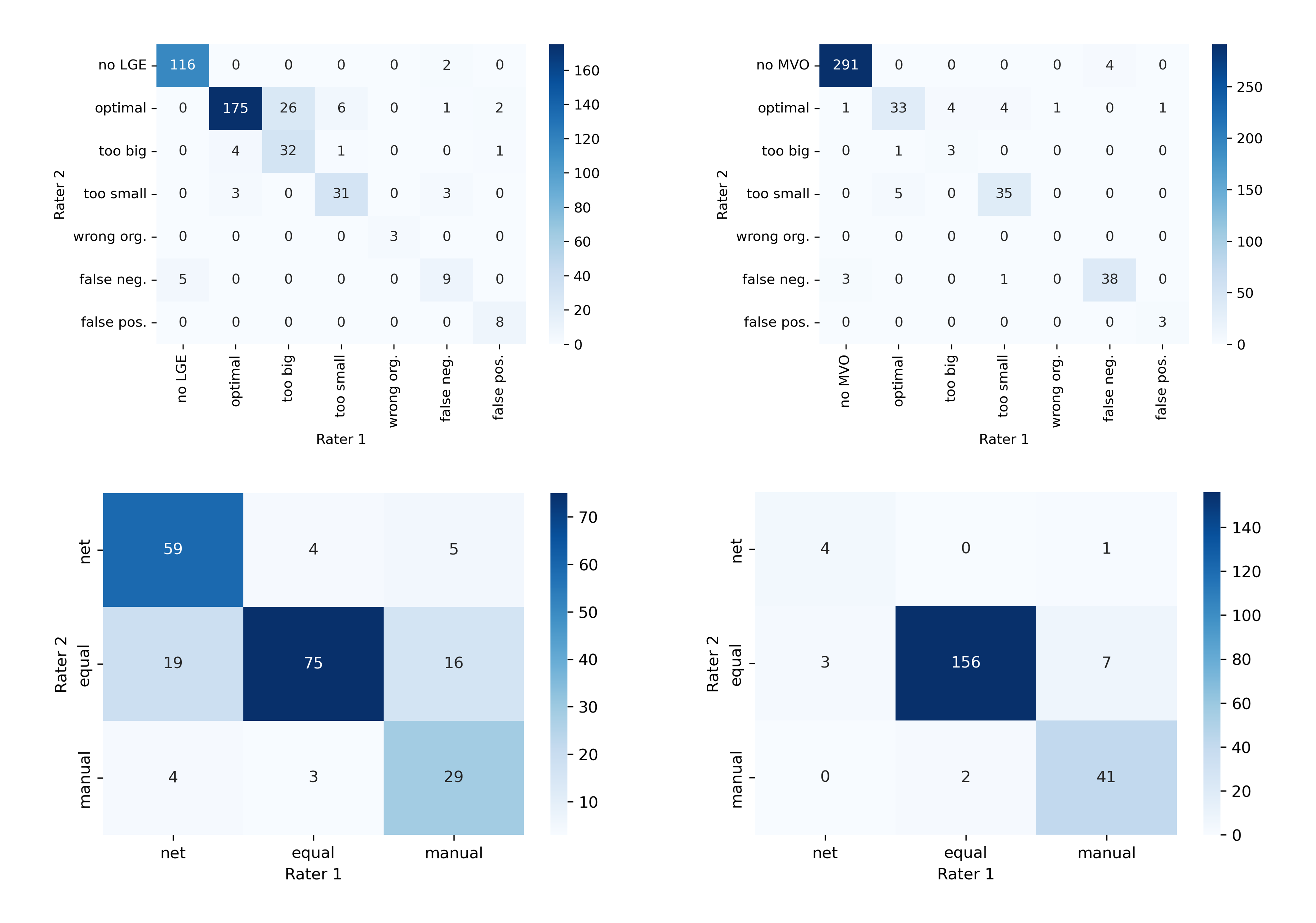}
    \caption{Confusion matrices on rater agreements for single image evaluation (top) and method comparison (bottom) for LGE (left) and MVO (right) segmentations.}
    \label{fig:ratings}
\end{figure}

In addition, to confirm the informative value of the subjective ratings of the two experts, we evaluated their agreement on a subset consisting of $20$ patients (Figure \ref{fig:ratings}). Calculating Cohen’s kappa coefficients ($\kappa$) revealed that the strength of agreement between the raters was very good for both LGE ($\kappa = 0.82$) and MVO ($\kappa = 0.88$) ratings. In rating LGE segmentation, the biggest difference between the experts was that in $26$ out of the total $212$ slices, rater 1 decided for the LGE segmentation to be too big, while rater 2 considered it as optimal. Similar to that, for MVO assessments in $4$ cases each, rater 1 was of the opinion that the markings were too large or too small, while rater 2 opted for optimal. In contrast to that, in five images, rater 1 voted for an optimal segmentation, while rater 2 assessed the MVO segmentation to be too small. When answering the question which of the segmentation was better, calculating linearly weighted Cohen’s kappa revealed good agreement ($\kappa = 0.64$) and very good agreement ($\kappa = 0.83$) between the raters for LGE and MVO segmentation, respectively. Especially for LGE segmentation, we could observe the trend that rater 1 decided more often for the two segmentations to be equally good, whereas rater 2 still decided for one of the segmentations.

\section{Discussion}

In this work, we developed and evaluated a deep learning-based pipeline that allows to quantify infarct scars and MVO from LGE CMR images in a fully-automated way. Compared to existing methods, our framework allows to perform the segmentation of infarcted areas without any manual preprocessing steps. Although there are public datasets for LGE CMR images available \cite{emidec_dataset,zhuang2018multivariate}, they do not reflect clinical reality. For instance, these datasets consist of image stacks that only include slices where the myocardium of the LV is visible. Also, often images were preprocessed such that the LV is located in the center of each image. As this cannot be expected when dealing with data in a clinical setting, the methods developed on these public datasets would rely on suitable preparation of the raw data before they can be applied. In contrast to that, in our framework, these preprocessing steps are also automated, making them easily applicable in daily practice. This complete automation from raw MR data to the final clinical markers is a huge time saver. On a conventional clinical computer without any GPU assistance (Intel(R) Core(TM) i7-10700 CPU @ 2.90GHz), our framework required between $3$ and $5$ seconds per patient ($3.9 \pm 0.4$). Compared to the manual measurements, which were reported to take around $20$ minutes per patient, this is a huge improvement. In contrast to prior work on automated infarct quantification, we evaluate the performance of our framework on a significantly larger dataset. Although a multi-center, multi-vendor study with a test dataset of $207$ patients was done for automated myocardial scar quantification in hypertrophic cardiomyopathy \cite{fahmy2020three}, to the best of our knowledge, no such study on this scale has been performed for STEMI patients. In our study, we tested the proposed method on data from $121$ unique STEMI patients, which is a big increase compared to previous studies that have only evaluated their methods using data sets up to $50$ patients. In addition, previous work only evaluated segmentation quality quantitatively by calculating metrics such as Dice coefficients, Hausdorff distances, or volume differences compared to human-generated ground truth measurements. We are the first to also perform a qualitative evaluation of automated myocardial scar segmentation, which provides additional information about the types of segmentation errors made by neural networks but also by humans. Here our results show that automated infarct segmentation exceeds manual measurements performed in clinical practice. However, it also shows that for detection and labeling of MVO, human raters still perform significantly better than the AI framework, which implies that there is still some potential for improvement in this area. Finally, we examine rater agreement of the subjective quality measures on a small subset of our data. This differs from previous work in that only rater agreement has been reported for quantitative metrics in myocardial infarction segmentation.  

The deep learning segmentation accuracy for myocardial infarction was very high. Of all the $152$ examinations that we evaluated, only two minimal scars were entirely missed by the framework. Also, when comparing to manual segmentations, the mean Dice score of $64.11 \%$ for myocardial infarction is comparable to reported inter-observer Dice scores of $56.9 \%$ \cite{li2023myops} and $69\%$ \cite{emidec_dataset} on other LGE CMR datasets. Dice coefficients for MVO segmentation were even higher than for LGE segmentation. However this was mainly due to the high specificity of our framework combined with the high number of patients ($129/152$) in our test dataset, which had no MVO. Bland-Altman analysis of scar percentages within the myocardium showed a low bias of $-1.26 \%$ and limits of agreement ranging from $-11.57$ to $9.05 \%$, which was considered to be quite high by experienced physicians. In comparison, \cite{fadil2021deep} report a mean difference of $4.4 \%$ and limits of agreement from $-10.6$ to $19.4 \%$ for intra-observer scar measurements, which confirms that the accuracy of our automated infarct size measurements is comparable to human measurements. Also, AVD and AVDR were comparable to values obtained in different studies, such as those reported on the EMIDEC Challenge leaderboard (\url{https://emidec.com/leaderboard}).

Our qualitative evaluation of the segmentation performance revealed that, especially for LGE segmentation, the deep learning framework was able to outperform human readers in terms of subjective segmentation accuracy. In summary, two CMR experts considered the AI-created segmentation masks to be optimal more often than the human-created ones. Also, they preferred the deep learning measurements significantly more often when comparing directly. Interestingly, AI also had a better sensitivity, as fewer myocardial scars were overlooked compared to humans on a per-slice level. This shows that human measurements can also be prone to errors, as even trained observers can overlook small myocardial scars in individual slices, presumably due to time pressure or lack of concentration. On the other hand, for detection and quantification of MVO, it turned out that manually created measurements were still superior. Although specificity between humans and AI was comparable, there was a big difference in sensitivity as a lot of MVOs were overlooked or only marked partly by the deep-learning-based framework. The reason for this could be that MVO, if present, only covers very small areas and additionally has a very similar pixel intensity to healthy myocardium. Additionally, in the training dataset of our framework MVO, was present in only $36$ out of the $144$ patients, further increasing the class imbalance. Therefore, we think that adding more patients with MVO to the dataset and also better addressing the class imbalance for MVO during training could improve the performance of the AI framework. However, as also human readers missed a significant portion of the present MVO, it is clear that MVO quantification is a very complex task requiring both medical knowledge and experience.

Although an underestimation of MVO size on LGE images compared to first-pass perfusion or early-gadolinium enhancement sequences is known \cite{mather2009appearance}, the images were identical for human observers and the CNN, so the results should still be comparable. Furthermore, first-pass perfusion images have lower signal-to-noise ratio, spatial coverage, and ventricular coverage, whereas LGE imaging has high spatial and contrast resolution \cite{kim1999relationship} and enables full coverage of the LV myocardium. However, this should facilitate the fundamental detectability of MVO in LGE images for both humans and the proposed network, thus allowing for a fair comparison.

In our study, we found very good agreement between the subjective quality assessments between the two raters for both MVO and LGE segmentations. This confirms the relevance of our qualitative study and suggests that further research should be carried out in this area, as articles that include clinical experts’ evaluation of medial segmentation quality are generally rare \cite{hoebel2023expert}.

Our framework had some limitations. First, our study was only concerned with data coming from one single hospital, and secondly, all the images were acquired at 1.5T with scanners from one single vendor. Thus, this study cannot demonstrate that our framework will work across different scanners or potentially different MR protocols. Another challenge compared to publicly available LGE datasets \cite{li2023myops,emidec_dataset} is that on the test data set, the manual segmentations were not all done at the same time but over the years by different members of the research working group. Also, when considering the results of the qualitative evaluation, one could argue that even the ground truth we used for training has errors, which would mark a clear limitation. However, we suggest that this could also be an inherent problem of infarct segmentation in general, since sometimes the boundaries between healthy and infarcted tissue are not clear-cut and leave room for interpretation. Investigating this by qualitatively analyzing ground truth segmentations of other LGE datasets could be of interest for future work.  Finally, there were no CMR images of healthy patients in any of the data we used, which is a serious limitation compared to other LGE datasets, such as the EMIDEC dataset, where $33 \%$ of all MR examinations are of healthy patients. This does not allow us to make a statement about whether our framework might erroneously mark infarct scars in examinations of healthy patients.

In conclusion, our fully automated infarct segmentation pipeline is able to compete with human experts. Although there are still some weaknesses in MVO segmentation, we have shown that our infarct segmentation algorithm outperforms trained human observers in qualitative segmentation accuracy. This, together with the associated massive time savings, paves the way for a potential application of fully automated CMR myocardial infarction quantification in clinical practice.

\section*{Autors contributions}
\textbf{Matthias Schwab:} Methodology, Software, Writing - Original Draft, Conceptualization. \textbf{Mathias Pamminger:} Writing - Review \& Editing, Investigation. \textbf{Christian Kremser:} Conceptualization, Writing - Review \& Editing. \textbf{Markus Haltmeier:} Conceptualization, Supervision, Writing - Review \& Editing. \textbf{Agnes Mayr:} Project administration, Supervision, Writing - Review \& Editing.

\section*{Disclosures of conflicts of interest}
This work was supported by the Austrian Science Fund (FWF) [grant number DOC 110].

\section*{Supplemental Material}
\label{supplemental}

\appendix
\renewcommand{\thesection}{S\arabic{section}} 
\setcounter{section}{0} 
\renewcommand\thefigure{S\arabic{figure}}  
\setcounter{figure}{0} 
\renewcommand{\theHfigure}{S\arabic{figure}} 

\captionsetup[figure]{name=Fig}

\section{Dataset Specifics and Imaging Protocols}
\label{sec:S1}

For all the patients used in our study, inclusion criteria were: first STEMI according to the ESC/ACC committee criteria \cite{thygesen2018fourth}, revascularization by p-PCI within $24$ hours after onset of ischemic symptoms, and Killip class $<3$ at time of CMR. The following exclusion criteria were applied: age $<18$ years, TIMI flow 0 and 1 after p-PCI, any history of previous myocardial infarction or a non-fatal reinfarction during the study period, an estimated glomerular filtration rate $<30$ ml/min per $1.73$ m², and any other contraindication to CMR examination. 
Each patient underwent LGE CMR imaging after a minimum interval of $10$ minutes after intravenous application of a gadolinium bolus of $0.2$ mmol/kg body weight (Gadubutrol, Gadovist, Schering, Berlin). This procedure employed an ECG-triggered phase-sensitive inversion recovery (PSIR) single-shot TrueFISP sequence with consecutive short axis slices. Parameters for this scan included a slice thickness of $8$ mm, an interslice gap of $2$ mm, a FOV of $400 \times 363$ mm, voxel dimensions of $2.2 \times 1.6 \times 8.0$ mm, a TR of $590$ msec, a TE of $1.2$ msec, a flip angle of $45^{\circ}$, and a GRAPPA iPat factor of $2$.

\section{Manual Segmentation}
\label{sec:S2}

\begin{figure}[h!]
    \centering
    \includegraphics[width=\textwidth]{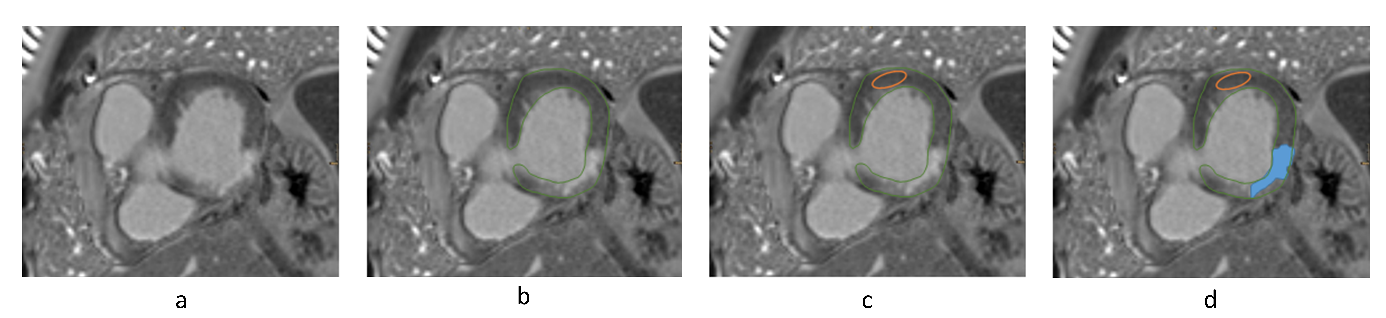}
    \caption{Manual segmentation procedure. Slice is checked for LGE-enhanced tissue (a). If infarction is present, epi and endocardial borders (green) are drawn (b). A ROI in the noninfarcted myocardial segment (orange) is drawn (c). After windowing using the 5SD method, the myocardial scar (blue) is marked (d). }
    \label{fig:S1}
\end{figure}

Semiautomated manual segmentations of the LGE images were performed using the local routine diagnostic interpretation and reporting software (DeepUnity Diagnost, Dedalus Healthcare Systems Group, Germany). In a first step, the slices with visible myocardial scars were selected manually. On these slices, epi and endocardial borders were outlined. Then within the myocardium, a region of interest (ROI) was drawn in the non-infarcted myocardial segment opposite to the scar to define the remote myocardium (Fig \ref{fig:S1}). From the mean signal intensity and standard deviation (SD) of this ROI, a threshold of 5 SDs above the mean signal was chosen. This threshold is based on the literature and our own experience \cite{mayr2022evolution} and its goal is to avoid biased assessment of hyperenhanced myocardial regions due to subjective window settings. After windowing with this threshold, the contours of the LGE-enhanced areas were drawn manually. In this way, infarct area was assessed quantitatively for each slice and segment. Infarct volume in ml was then calculated by multiplying the hyperenhanced area with slice thickness, including the interslice gap. MVO volumes were calculated the same way, but unlike for LGE segmentations, MVO was labeled on the original images without the use of windowing.  

Before performing segmentations on the actual datasets, all human raters were trained on a separate dataset consisting of $30$ patients (BL, 4FU, and 12FU examinations). Raters had to achieve acceptable inter-rater reliability before they were allowed to perform segmentations on the actual datasets.

\clearpage

\begin{figure}[!b]
    \centering
    \includegraphics[width=.9\textwidth]{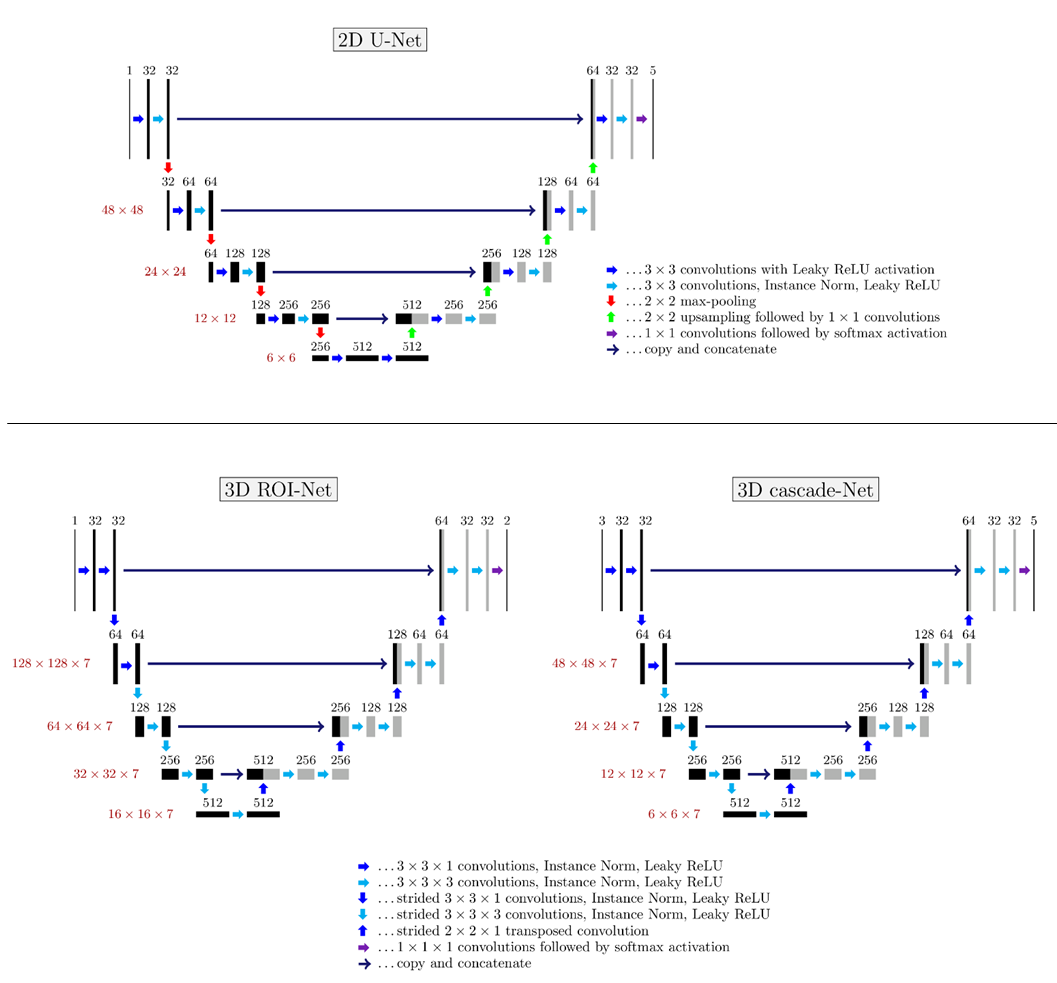}
    \caption{Manual segmentation procedure. Slice is checked for LGE-enhanced tissue (a). If infarction is present, epi and endocardial borders (green) are drawn (b). A ROI in the noninfarcted myocardial segment (orange) is drawn (c). After windowing using the 5SD method, the myocardial scar (blue) is marked (d). }
    \label{fig:S2}
\end{figure}

\section{Network Architectures and Training Details}
\label{sec:S3}

For our framework, we trained a total of 3 different networks: 

\begin{enumerate}
    \item A 3D U-net (ROI-net), which is trained to segment the left ventricle in the original clinical MR images.
    \item A 2D U-net that is trained to do multiclass segmentation of the left ventricle on a per-slice level.
    \item An error-correcting 3D U-Net that improves the 2D segmentation masks for infarction and MVO by including information about neighboring slices.
\end{enumerate}

All the networks were trained on the same training dataset. For the ROI-net, we used the manually marked epicardial boundaries to create binary segmentation masks of LV and background. Because the original image stacks have different image sizes, we cropped or zero-padded the images to a uniform size of $256 \times 256$. 

For all networks, we use basic U-Net architectures with kernel sizes of $3 \times 3$ for the 2D network and $3 \times 3 \times 3$ for the 3D networks, respectively (Fig \ref{fig:S2}). After each convolutional block instance normalization is applied, followed by a leaky rectified linear unit (ReLU) activation function. For the 2D U-Net, downsampling is achieved by max-pooling and upsampling by bilinear interpolation. The 3D CNNs use strided convolutions, following \cite{szegedy2016rethinking}, for downsampling as well as strided convolution transposed for upsampling. For the 3D U-Nets, no downsampling or upsampling is performed with respect to the third dimension. ROI-net and the 2D-net have only one input channel. For the final error-correcting 3D CNN, the number of input channels is $3$, as the course 2D segmentations of myocardial scar and MVO are added as additional information to the input of the CNN. 

\begin{figure}[!b]
    \centering
    \includegraphics[width=.7\textwidth]{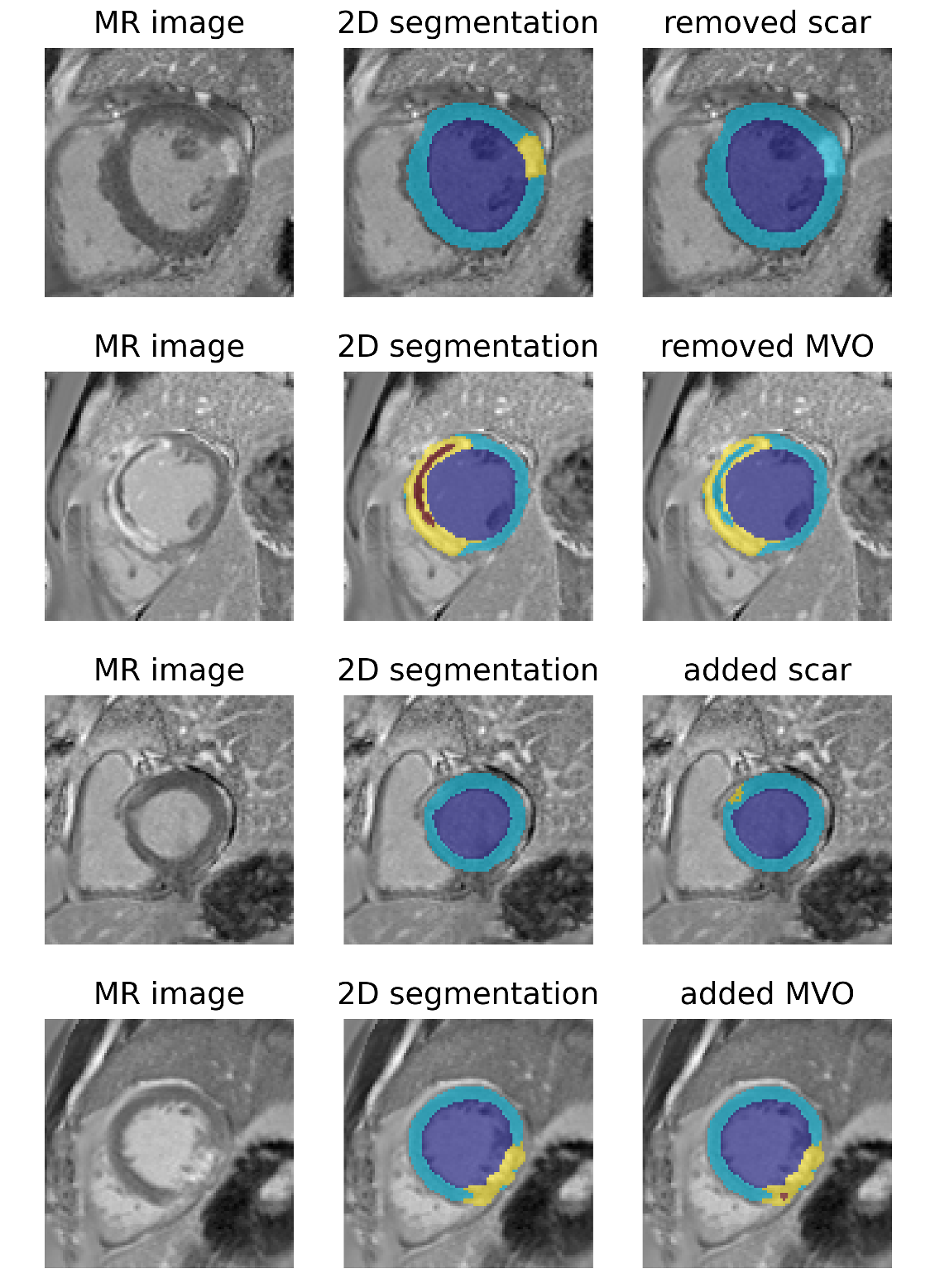}
    \caption{Examples for artificial modifications of the 2D segmentation masks by the perturbation module. The first two lines show examples where correctly identified infarction resp. MVO was removed. Lines 3 and 4 show images where the perturbation module added an artificial scar resp. MVO.}
    \label{fig:S3}
\end{figure}

ROI-net was trained for $300$ epochs on a batch size of $4$, and 2D and 3D cascade nets were trained for $750$ epochs with batch sizes of $32$ and $4$, respectively. All models were trained from scratch with randomly initialized network weights. A variation of the Dice loss as proposed in \cite{milletari2016v} was used as the loss function for all the networks. The loss function was minimized through stochastic gradient descent with Nesterov momentum ($\mu = 0.99$) and an exponentially decaying learning rate. Before AI input, all images were normalized to have a zero mean and a standard deviation of one. Segmentation performance was evaluated after each training epoch on $33$ CMR images on an additional evaluation dataset, which was obtained the same way as the training dataset, and the best-performing model was finally chosen. During training, different data augmentation techniques were applied (Gaussian blurring, gamma correction, additive white noise, changing contrast/brightness, simulating lower resolution, translation, flipping, elastic deformations, scaling). For the cascaded CNN, the perturbation module was switched on only after $100$ epochs of training. Further, deep supervision \cite{lee2015deeply} was used, additionally providing direct supervision to some hidden layers instead of only supervising the output layer. All the training of the framework was performed on a NVIDIA A40 GPU using the Pytorch deep learning library.

\section{Perturbation Module}
\label{sec:S4}

In our training pipeline, we use an error-correcting 2D-3D cascaded CNN pipeline, which was especially created infarct segmentation and was initially proposed in \cite{schwab2023error}. In this framework, first a 2D CNN performs multiclass segmentation on a per-slice level. Then, a 3D CNN corrects the coarse 2D segmentation masks by incorporating information about neighboring slices. However, since the 2D CNN was optimized on the training dataset, its segmentation masks are already quite precise on this data. This fact does not really allow the subsequent 3D segmentation network to learn to detect and improve 2D segmentation errors during training. The perturbation module, which is interposed between the 2D and 3D networks, addresses this issue of overly precise coarse 2D segmentation masks by incorporating 2D-characteristic errors. More specifically, the module is designed to simulate potential segmentation errors that the first-stage network might encounter due to limited inter-slice information. By artificially generating these errors during training, the subsequent 3D network is enforced to learn how to correct them effectively. The perturbation module introduces several types of artificial segmentation errors (Fig \ref{fig:S3}):

\begin{itemize}
    \item \textbf{Enhanced data augmentation:} The ranges for contrast, brightness, low resolution, and gamma augmentations are increased beyond what is used in the 2D augmentation pipeline. 
    \item \textbf{Random class deletion:} For certain classes (myocardial scar, MVO, or both), the 2D segmentations are deleted in one single slice of  the volume.
    \item \textbf{Infarction misinformation:} Incorrect infarction annotations are added to random slices. This is done by identifying the 85th percentile of pixel values within the myocardium on a chosen slice and adding the largest connected component as a false scar annotation.
    \item \textbf{MVO misinformation:} MVO annotations are falsified by changing the label of a pixel classified as scar, along with some neighboring pixels, to MVO.
    \item \textbf{Complete mask nullification:} In some cases, the entire 2D segmentation mask is set to zero for random samples.
\end{itemize}

During training, these perturbations are applied with low probabilities ($10\%$ for deleting single classes, nullifying the mask, or adding false infarction annotations, and $2\%$ for adding incorrect MVO annotations). Since these perturbations are typically applied to random single slices within a 3D volume, the 3D network is trained to detect and correct errors characteristic of missing inter-slice information, improving the generalizability properties of the final framework on unseen data.

\section{Qualitative Evaluation}
\label{sec:S5}

\begin{figure}[t]
    \centering
    \includegraphics[width=\textwidth]{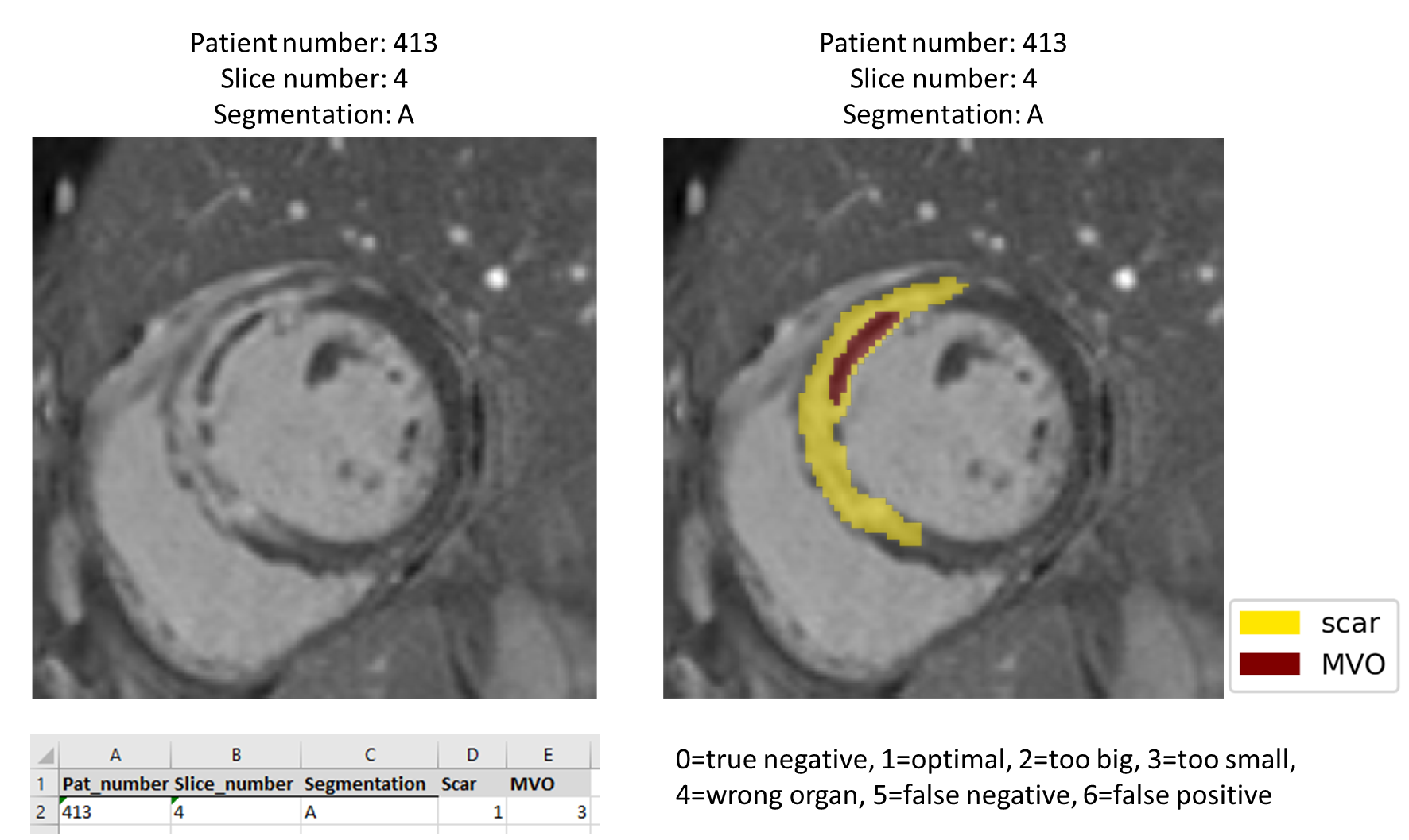}
    \caption{Rating process on the individual images. Slice number and anonymized patient number were displayed to the raters. For each image, they could choose between seven different subjective rating categories for both scar and MVO segmentation.}
    \label{fig:S4}
\end{figure}

\begin{figure}[h!]
    \centering
    \includegraphics[width=\textwidth]{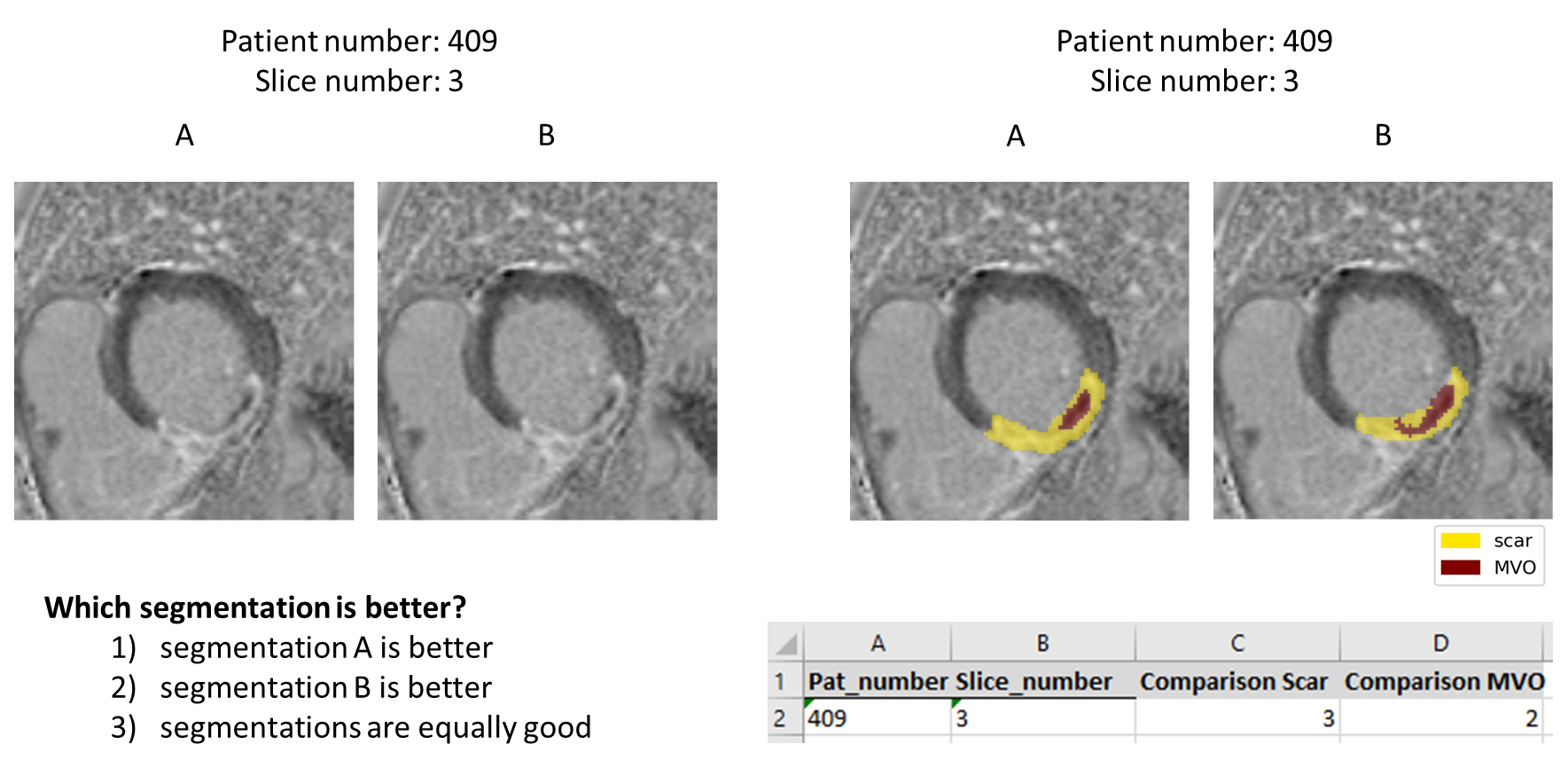}
    \caption{Subjective comparison of human and AI-based segmentations. Not knowing which segmentation belongs to which method, the raters compared the two segmentations and reported their assessment.}
    \label{fig:S5}
\end{figure}

Qualitative evaluation was done on a per-slice level. For each patient, the medical raters could click through all the CMR slices and for each slice they individually assessed subjective segmentation quality. They were able to click back and forth between the raw input image and the segmented image. Without knowing if the segmentation was done manually or automatically, they had to decide for one of the seven segmentation ratings (Fig \ref{fig:S4}) and type their assessment into an Excel table. 
For the comparison between the two methods, both segmentations were presented to the raters side by side. Again, by clicking back and forth, they could display and hide the segmentation masks (Fig \ref{fig:S5}). Manual and automatic segmentations were distributed randomly between A and B such that the raters had no information about which segmentation was done by which method. Comparing the two images, the raters had to decide if they prefer one of the segmentations or if they consider them equally good. Again, the rating was carried out for both classes (MVO and scar) and the evaluations were inserted into an Excel sheet.  

\clearpage

\clearpage
\begin{figure}[t]
    \centering
    \includegraphics[width=\textwidth]{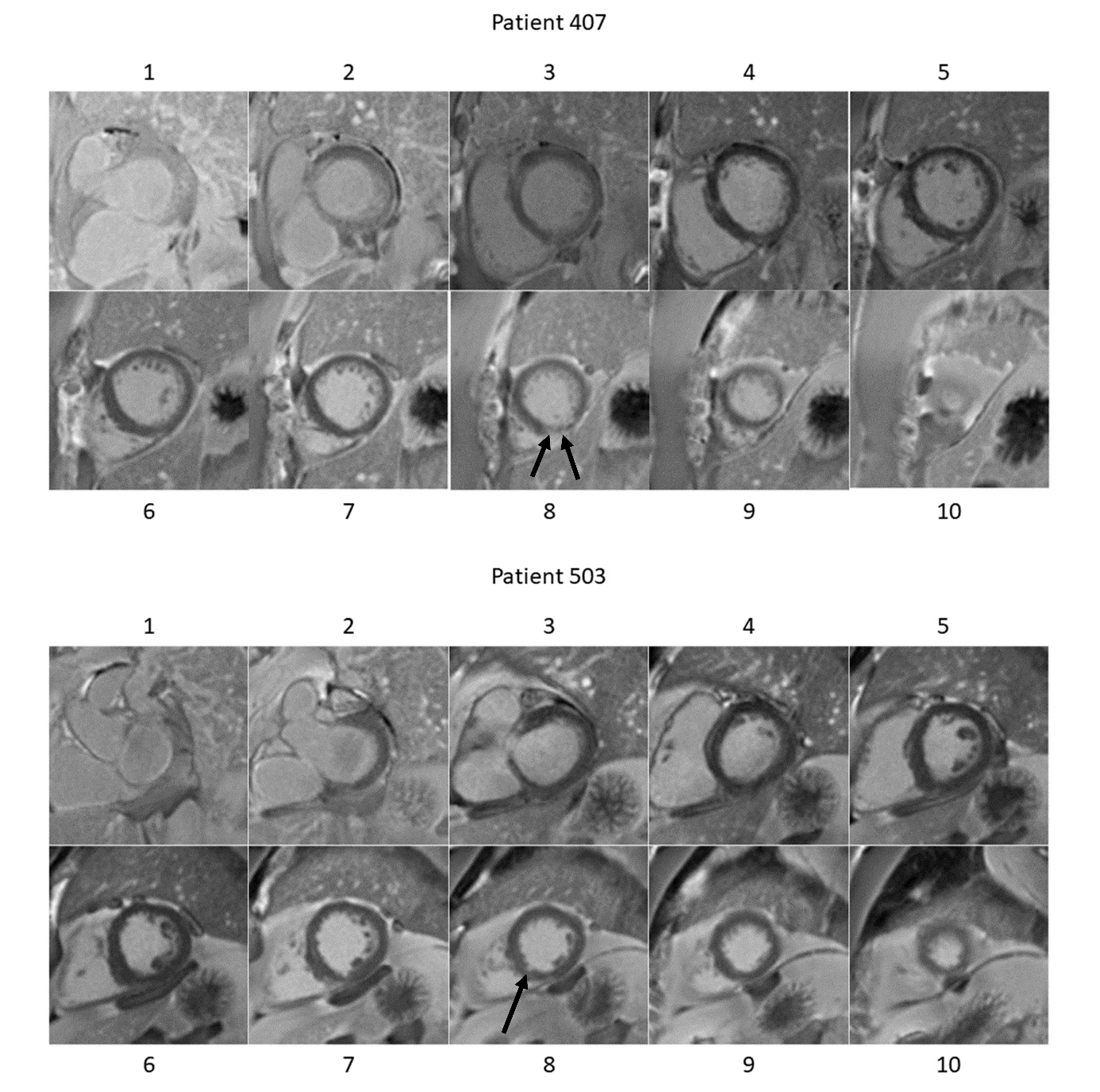}
    \caption{The two patients in whom the proposed method missed the scars. Patient 407 has a very small posterior wall infarct scar apically (Slice 8). Patient 503 has a very hard-to-see scar in slice 8. Both scars are almost impossible to detect in the LGE images and could only be clearly confirmed after additional reviews of the functional images and previous examinations.}
    \label{fig:S6}
\end{figure}

\bibliographystyle{vancouver}  
\bibliography{references}  
\end{document}